# Nano-Objects Developing at Graphene/Silicon Carbide Interface

S. Vizzini,[1] H. Enriquez,[1] S. Chiang,[1,2] H. Oughaddou [1,3] and P. Soukiassian [1]

1 - Commissariat à l'Energie Atomique, Laboratoire SIMA, DSM-IRAMIS-SPCSI, Bât. 462, Saclay, 91191 Gif sur Yvette, and Département de Physique, Université de Paris-Sud, 91405 Orsay, France

2 - Department of Physics, University of California-Davis, CA 95616-8677, U.S.A.

3 - Département de Physique, Université de Cergy-Pontoise, 95031 Cergy-Pontoise, France

Abstract

We use scanning tunneling microscopy and spectroscopy to study epitaxial graphene grown on a C-face 4H-SiC(000-1) substrate. The results reveal amazing nano-objects at the graphene/SiC interface leading to electronic interface states. Their height profiles suggest that these objects are made of packed carbon nanotubes confined vertically and forming mesas at the SiC surface. We also find nano-cracks covered by the graphene layer that, surprisingly, is not broken, with no electronic interface state. Therefore, unlike the above nano-objects, these cracks should not affect the carrier mobility.

Graphene is a single sheet of graphite studied by theory half a century ago and then epitaxially grown on substrates or exfoliated [1-4]. Recently, it was shown to have exceptional transport properties with charge carriers moving at $\approx$ zero mass and constant velocity just like photons [5,6]. The epitaxial growth of graphene on a silicon carbide (SiC) substrate is of special interest. SiC is a wide-band-gap IV-IV compound semiconductor with gap ranging from 2.4 eV to 3.3 eV depending on the polytype [7-9]. Therefore, epitaxial graphene on SiC is especially promising for applications that are significant for electronics of the future [2,4,5]. Indeed, it has recently been added to the Roadmap of Semiconductor Technology. Such epitaxial growth on the 6H/4H-SiC(0001) Si-face and on the 6H/4H-SiC(000-1) C-face has been studied extensively, leading to the determination of atomic/electronic structures and transport properties, using advanced experimental techniques such as atom-resolved scanning tunneling microscopy and spectroscopy, core level and valence band photoemission spectroscopies, grazing incidence X-ray diffraction, and state-of-the-art theoretical calculations [4,6,10,11]. On the C-face, graphene multilayers can be grown epitaxially with each layer decoupled from one another [11], leading to unprecedented high carrier mobility up to 250,000 cm$^2$/V.s [6]. In contrast, for a single graphene layer epitaxially grown on the same C face 4H-SiC(000-1) substrate, the mobility is significantly lower at 20,000 cm$^2$/V.s [12]. This could be due to the quality of the graphene/SiC interface, probably as a result of the very harsh growth conditions (e.g. high temperatures) occuring during graphene formation, resulting in rapid Si depletion from the surface. Indeed, SiC interfaces are known to exhibit defects such as electronic interface and trap states e.g. at SiO$_2$/SiC interfaces [7] and crack formation [13]. Also, strain is the driving force in SiC surface ordering [8], leading to unprecedented self-organized nanostructures and surface transformations [7,14]. There had been no in-depth study to explore possible defects existing at graphene/SiC interfaces, especially at the atomic scale.

In this Letter, we use scanning tunneling microscopy and spectroscopy (STM/STS) to study the graphene/SiC interfacial region. We find amazing nano-objects that are laterally confined by and below the graphene layer on top of the SiC surface. These nano-objects, which are likely to be carbon nanotubes (CNT), cause the formation of specific electronic interface states, possibly resulting in detrimental effects on the graphene transport properties. We also find nano-crack defects on the SiC surface that are covered by the graphene layer with no modification of the STS spectral response and no resulting specific electronic interface state, unlike the nano-object behavior.

The STM/STS experiments were performed using a variable-temperature Omicron instrument at pressures in the low $10^{-10}$ torr range. Graphene was grown on top of a C-face 6H-SiC(000-1) n-doped at $10^{-16}$/cm$^3$ substrate (Cree), by annealing between 1200°C and 1300°C, similar to well-established procedures [10,15,16]. Single and double graphene layers are identified from characteristic honeycomb and atomic structures seen in STM topographic images [10,15,16].

Fig. 1 displays representative STM topographs of single graphene layer grown on a stepped SiC surface by annealing at 1300°C for 2 minutes. The 420x200 nm$^2$ area in the center of Fig. 1a covered by a network of bright nano-objects, is of special interest. Such features appear occasionally on other areas of the surface. Further details about these nano-objects are found in a higher resolution image (Fig. 1b), with a three dimensional (3D) topograph (Fig. 1c) revealing their shape. A representative height profile along XX' (Fig. 1d) reveals that these amazing nano-objects have a constant height of about 7Å, extremely steep sides, and very flat tops. The 3D view in Fig. 1c better emphasizes these striking characteristics, showing that these nano-objects form mesas all having the same height on top of the SiC surface.

In order to identify whether or not these nano-objects are covered by graphene, we now change the tunneling conditions. In Fig. 2a and 2a', we observe the characteristic honeycomb structure of a single graphene layer. The later covers not only the nanostructures but also the whole surface as a continuous single atomic layer. In comparison, Fig. 2b and 2b' provide a similar view of a representative area that does not have such nano-objects. Fig. 2b shows a representative topograph of a graphene sheet covering the SiC surface, which also exhibits the well established honeycomb structure of a single graphene layer together with the Moiré characteristic of the coupling with the substrate (Fig. 2b') [15,16].

Graphene layer growth results from Si sublimation leaving excess carbon species that bond as *sp2* to form graphene. Due to the harsh growth conditions using elevated temperatures (1300°C in this study), the species that could be trapped below the graphene layer could either be silicon, or carbon that did not form graphene. The striking square cross-section shape of these nano-objects clearly suggests that they are due to CNT growing vertically during graphene formation, remaining trapped below the graphene sheet. On the other hand, Si species trapped below the graphene layer would most likely form two-dimensional (2D) atomic layers. Even if 3D Si objects are considered, they are most unlikely to exhibit the obserevd shape with sharp vertical sides. Indeed, CNT are known to grow perpendicularly to the surface of 4H-SiC(000-1) substrates [17], which further supports the interpretation of these nano-objects as CNT grown below the graphene layer.

In order to get deeper insights about these nanostructures, we now perform STS measurements for the single layer graphene covering the nano-objects and the SiC surface. Fig. 2c gives the measured I(V) curves showing the Schottky barrier character of the graphene layer for both monolayer graphene on SiC (black squares) and on top of a nano-object (red circles). However, the results also indicate a significant difference in the slope for positive bias, suggesting that these nano-objects are likely to induce new unoccupied electronic states at the graphene/SiC interface.

To better understand this aspect, we look at Fig. 2c' which displays the (dI/dV)/(I/V) derivative curve of the same representative area as in Fig. 2c, providing the local density of states for both graphene-covered SiC and nano-objects. We can clearly detect major differences between these curves, with the graphene-covered nano-objects exhibiting three new spectral features, two in the valence band (IS 1,IS 2) and one in the conduction band (IS 3). IS 3 is located at 100 meV above $E_F$, while IS 2 and IS 1 are located at 85 meV and 150 meV respectively below $E_F$. Note that the only significant spectral feature for the graphene-covered SiC is peak M, located at 120 meV above $E_F$. IS 3 has a very different line shape from peak M, including a much larger intensity and a full-width-at-half-maximum (FWHM) of $\approx$ 80 meV, i.e. about 30% smaller than peak M. Therefore, IS 3 is unlikely derived from peak M and has a different character due to the underlying nanostructure. It is an empty electronic interface state, which may affect electron carrier mobility. On the other hand, filled interface states, IS 2 and IS 1 would influence hole carrier mobility.

The IS 3 feature in the conduction band suggests that its presence likely results from "open" rather than from "capped" CNT located below the graphene layer. In fact, the C atoms located at the top of an open CNT, just below the graphene layer, would have empty dangling bonds influencing the graphene layer just above, possibly leading to electron depletion. Such a situation will not take place for a capped CNT (which would have its end-bonds all satisfied), suggesting that IS 3 results primarily from the interaction between uncapped CNT nano-objects and the graphene layer. In contrast, IS 1 and IS 2 in the valence band could possibly result from "capped" CNT, with filled electronic orbital overlap occurring at the graphene/capped CNT interface as the most plausible explanation. Since CNT grown vertically on a 4H-SiC(000-1) surface are generally capped [18], the present results suggest that, while a similar growth is taking place below a graphene layer, some of the CNT seem to have an open termination.

We now look at another type of nanostructure present at the graphene/SiC interface, cracks defects, known to develop at SiC surfaces [14]. Fig. 3a displays a 2D topograph of a representative area exhibiting cracks while Fig. 3b shows a 3D Fourier transform-filtered image of the same area showing everywhere the characteristic honeycomb pattern of a single graphene layer. Height profiles in various area show that the crack depth is constant at 1.5 Å as seen in Fig. 3c from a representative height profile along AA'. The crack is about 1.1 nm wide. We now explore whether this single graphene layer is disrupted by the crack as such a disruption could be very detrimental to the transport properties. Fig. 3b shows clear evidence that the graphene layer is not broken by the surface fracture, but instead goes down into the cracks (by 1.5 Å from Fig. 3c) forming a continuously uniform layer (Fig. 3b) despite the very harsh growth conditions. This behavior is consistent with the very high mechanical resistance of graphene [18]. Furthermore, molecular dynamic simulations on graphite fracture show that cracks could develop along the main cristallographic directions, i.e. along zig-zag or armchair [19]. Actually, the same situation occurs for cubic and hexagonal SiC [13]. Interestingly, an epitaxial graphene layer on SiC has its cristallographic directions rotated from those of the 4H-SiC(000-1) or 4H-SiC(0001) substrates [11]. Therefore, in addition to the very high mechanical resistance of graphene, this feature probably also explains why a crack developing on the SiC surface does not extend to the graphene layer.

Figures 4a and 4b display 2D and 3D STM topographs of another representative area exhibiting cracks but covered by a double graphene layer as shown from the characteristic hexagonal atomic pattern in the 6x6 nm$^2$ inset of Fig. 4a. The 3D image (Fig. 4b) shows that ripples occur at the crack edges. One can see a very long 1D crack (over 250 nm) along the diagonal of the images and an island of about 65x50 nm$^2$ surrounded by cracks near the top of the topograph (Fig. 4a & 4b). The representative height profile along XX' (Fig. 4c) shows the crack depth is 3Å, with ridges about 1 Å high on one side and 4 Å high on the other side, with the crack depth at 1.5 nm. These ridges likely result from graphene ripple formation at the crack edges, which could be caused by the very different expansion coefficients between graphene and SiC. The SiC temperature dilation coefficient is indeed 3 times larger compared to graphite [20] leading to compressive strain in the graphene layer [21]. This interesting aspect suggests these ripples to be formed when the substrate is cooling down after graphene formation at high temperature.

We next use STS to explore whether such crack defects on the SiC substrate would result in interface states as observed for the CNT nano-objects described above, despite the fact that graphene sheet is not broken. Fig. 4d displays (dI/dV)/(I/V) measurements for graphene covering a crack (top), and a crack-free surface (bottom). The two (dI/dV)/(I/V) curves are very similar with no specific electronic state showing above the cracks. Thus, unlike the above CNT nanostructures, these crack defects located below the graphene layer should have no detrimental effects on the carrier mobility.

In conclusion, our present results show evidence of two types of nanostructures at the graphene/SiC interface. These include nano-objects made of carbon nanotubes forming mesas with a constant height, which are laterally confined by the graphene layer. Their spectral response exhibits interface states that may potentially have detrimental effect on the transport properties. This finding is likely to be relevant, at least in part, to the lower carrier mobility of single epitaxial graphene layer compared to that of multiple graphene layers. In contrast, nano-cracks at the SiC surface do not affect the graphene layer which amazingly goes into the crack without breaking and with no resulting electronic interface states, suggesting that such cracks are unlikely to affect the graphene transport properties. This investigation illustrates a very interesting aspect of graphene, which is not only a highly resistant material, but also a highly pliable one, able to wrap a nano-object or to follow defects going deep inside a substrate fracture without disruption. It addresses one of the central issues common to semiconductor science and technology, namely the ability to understand and control the interface, especially for graphene where harsh conditions take place during growth.

Acknowledgements: This work is supported by ANR (Agence Nationale de la Recherche).

# References

1   I. Forbeaux, J.-M. Themlin and J.-M. Debever, Phys. Rev. B **58**, 16396 (1998).

2   C. Berger *et al.*, J. Phys. Chem. B **108**, 19912 (2004).

3   K.S. Novoselov *et al.*, Science **306**, 666 (2004).

4   T. Ohta, A. Bostwick, T. Seyller, K. Horn, and E. Rotenberg, Science **313**, 951 (2006).

5   C. Berger *et al.*, Science **312**, 1191 (2006).

6   M. Orlita *et al.*, Phys. Rev. Lett. **101**, 267601 (2008); D.L. Miller *et al.*, Science **324**, 924 (2009).

7   Silicon Carbide, A Review of Fundamental Questions and Applications to Current Device Technology, W.J. Choyke, H.M. Matsunami and G. Pensl editors, Akademie Verlag, Berlin, Vol. **I & II** (1998).

8   P. Soukiassian and H. Enriquez, J. Phys.: Cond. Mat. **16**, S1611-S1658 (2004).

9   K. Heinz, J. Bernhardt, J. Schardt and U. Starke, J. Phys.: Cond. Mat. **16**, S1705–S1720 (2004).

10  P. Mallet *et al.*, Phys. Rev. B **76**, 041403(R) (2007).

11 J. Hass *et al.*, Phys. Rev. Lett. **100**, 125504 (2008); M. Sprinkle *et al.*, Phys. Rev. Lett., *in press* (2009)

12 Xiaosong Wu *et al.*, arXiv: 0909.2903 (2009).

13 F. Amy, P. Soukiassian and C. Brylinski, Appl. Phys. Lett. **85**, 926 (2004); P. Soukiassian, F. Amy, C. Brylinski, T.O. Mentes and A. Locatelli, Mat. Sci. For. **556**, 481 (2007).

14 P. Soukiassian, F. Semond, A. Mayne and G. Dujardin, Phys. Rev. Lett. **79**, 2498 (1997); L. Douillard, V.Yu. Aristov, F. Semond and P. Soukiassian, Surf. Sci. Lett. **401**, L 395 (1998); V. Derycke, P. Soukiassian, A. Mayne, G. Dujardin, J. Gautier, Phys. Rev. Lett. **81**, 5868 (1998).

15 F. Hiebel, P. Mallet, F. Varchon, L. Magaud, J-Y. Veuillen, Phys. Rev. B **78**, 153412 (2008).

16 H. Yang *et al.*, Phys. Rev. B **78**, 041408R (2008).

17 M. Kusunoki, T. Suzuki, C. Honjo, H. Usami, H. Kato, J. Phys. D: Appl. Phys. **40**, 6278 (2007).

18 Changgu Lee, Xiaoding Wei, J.W. Kysar and J. Hone, Science **321**, 385 (2008).

19 A. Omeltchenko, J. Yu, R.K. Kalia and P. Vashishta, Phys. Rev. Lett. **78**, 2148 (1997).

20 D.K.L. Tsang, B.J. Marsden, S.L. Fok and G. Hall, Carbon **43**, 2902 (2005).

21 N. Ferralis, R. Maboudian and C. Carraro, Phys. Rev. Lett. **101**, 156801 (2008).

**Figure Captions**

Figure 1: a) STM topograph ($750 \times 1030$ nm$^2$) showing an area having self-organized nano-objects at the graphene/SiC interface (U= 0.5 V, I= 0.15 nA);

b) STM topograph ($143 \times 143$ nm$^2$) of nano-objects forming mesas below the graphene layer (U = 0.6 V, I = 0.2 nA); c) 3D view of the same as in b); d) Height profile along XX' showing the mesa nature of these nano-objects (at 7 Å high) with sharp vertical sides.

Figure 2: a) STM topograph ($87 \times 87$ nm$^2$) of nano-objects (U = -0.1 V, I = 0.2 nA); a') Detailed $7.2 \times 7.2$ nm$^2$ topograph showing the characteristic pattern of a single graphene layer covering a nano-object (U = -0.1 V, I = 0.2 nA);

b) STM topograph ($160 \times 160$ nm$^2$) of a single graphene layer covering a SiC surface (U = -0.6 V, I = 0.2 nA); b') Detailed $5 \times 5$ nm$^2$ area of b) showing the characteristic pattern of a single graphene layer covering SiC (U = -0.2 V, I = 0.1 nA);

c) STS I(V) characteristics of graphene on SiC (black squares) and graphene covering a nano-object (red dots); c') STS (d(I)/d(V))/(I/V) characteristics for the same area as in c).

Figure 3: a) STM topograph ($91 \times 58$ nm$^2$) of a single graphene layer covering an area having cracks (U = -0.1 V, I = 0.1 nA); b) 3D topograph (Fourier filtered transform) for the same area as in a); c) Height profile above a crack along XX'.

Figure 4: a) STM topograph ($194 \times 194$ nm$^2$) of a double graphene layer area covering cracks located at the SiC surface (U = 0.9 V, I = 0.2 nA) with the inset showing an area ($6 \times 6$ nm$^2$) with the characteristic Moiré pattern; b) 3D picture of the same area as in a); c) Height profile above a crack along XX' showing a 3Å depth; d) Bottom: STS (d(I)/d(V))/(I/V) characteristics on a double layer graphene on a flat SiC surface and d) Top: STS (d(I)/d(V))/(I/V) characteristics on a double layer graphene above cracks.

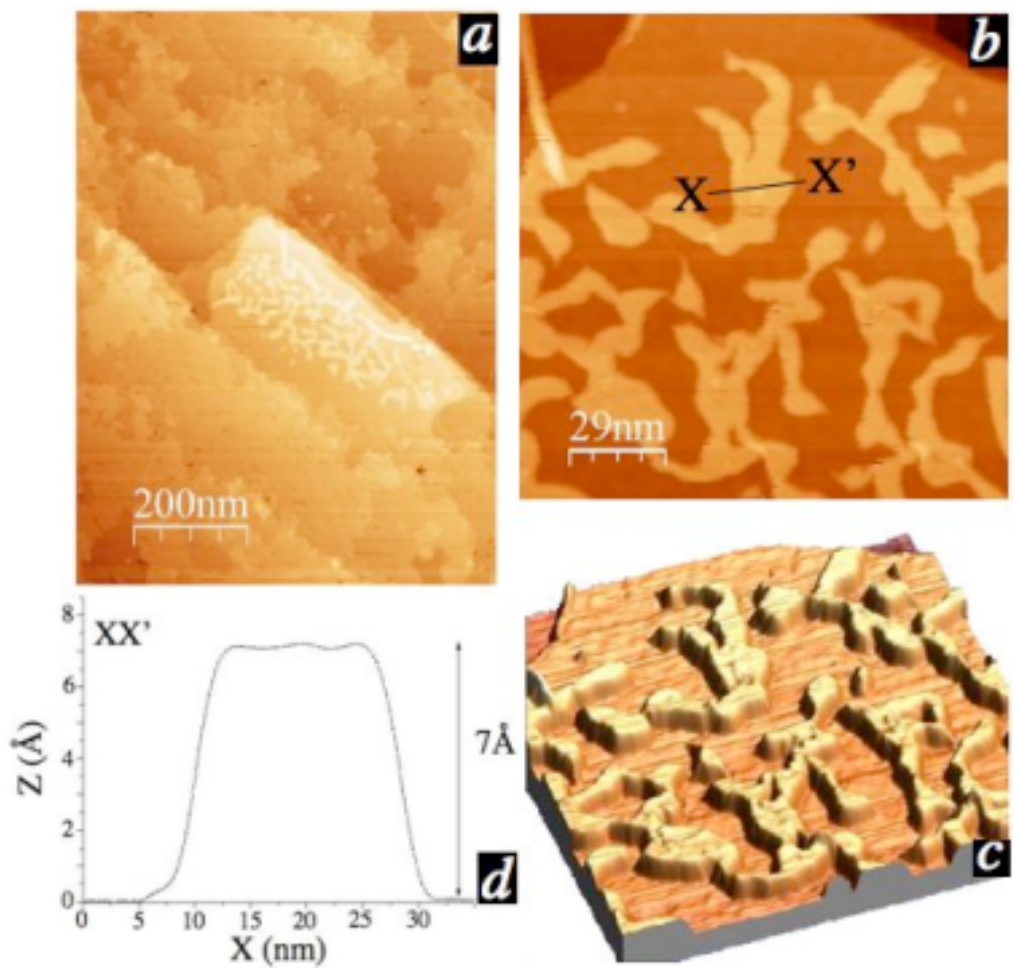

Fig. 1

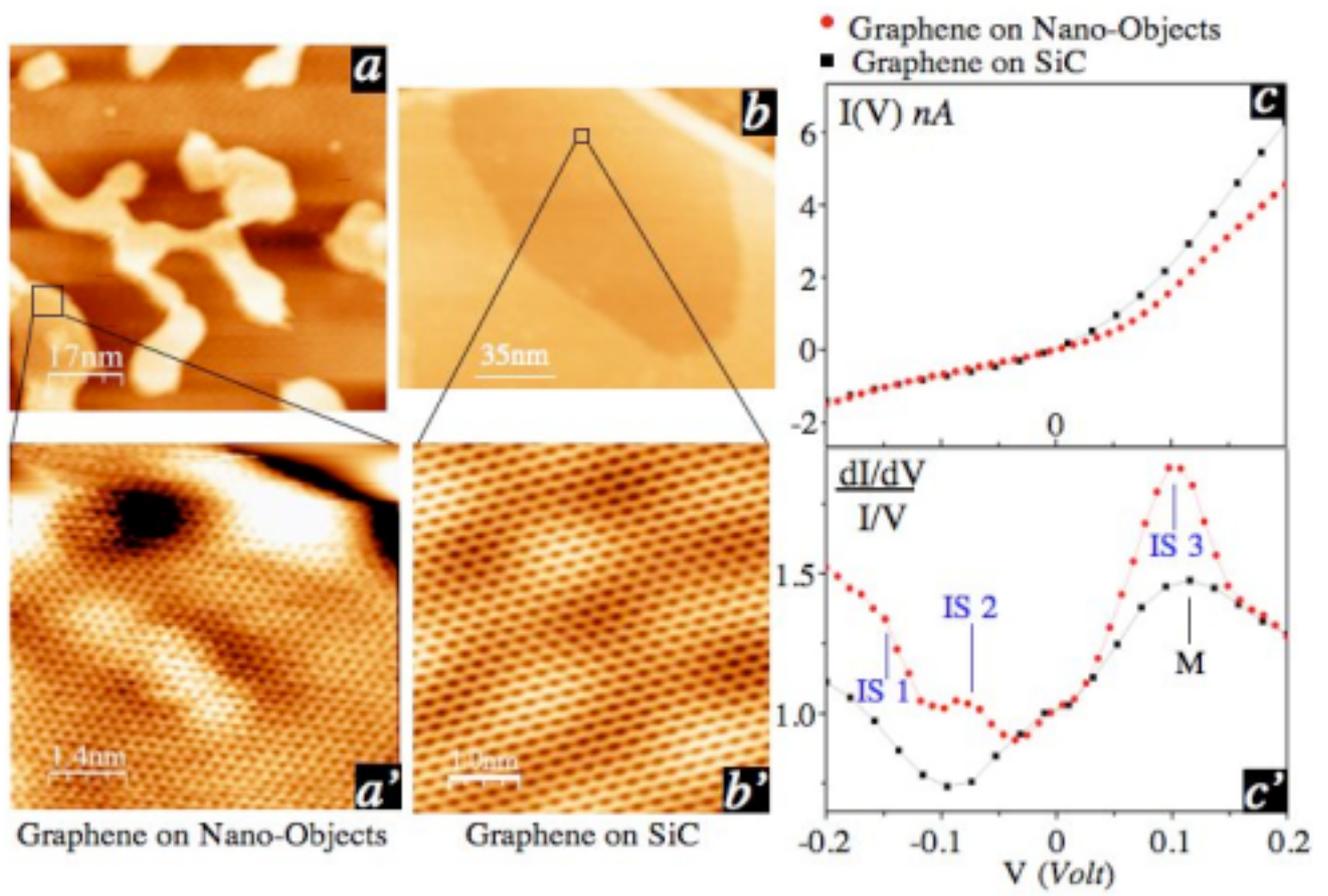

Graphene on Nano-Objects    Graphene on SiC

Fig. 2

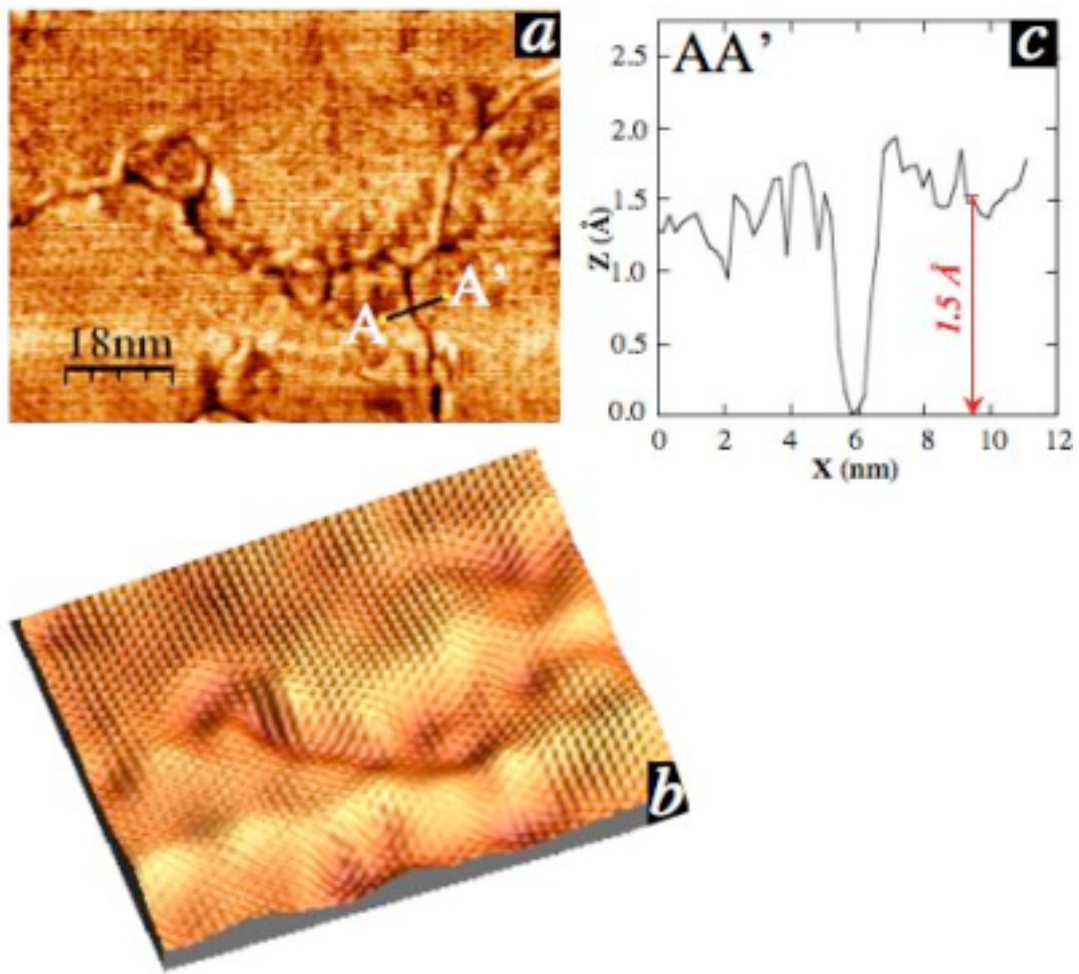

Fig. 3

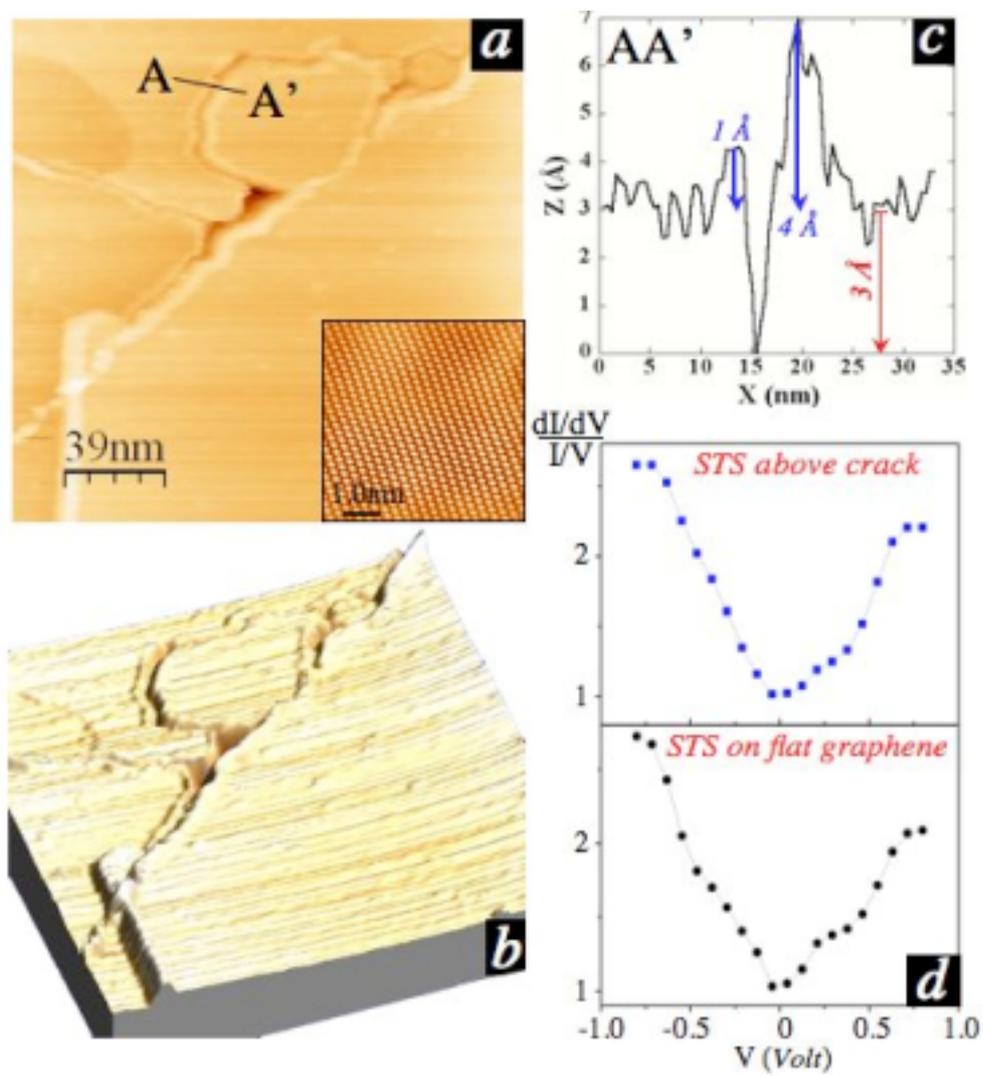

Fig. 4